%% file: acl2023.tex
\pdfoutput=1

\documentclass[11pt]{article}

\usepackage{ACL2023}

\usepackage{times}
\usepackage{latexsym}
\usepackage{lipsum}
\usepackage{subdef}
\usepackage{graphicx}
\usepackage{pgf}
\usepackage{algorithm}
\usepackage{algorithmic}[1]

\usepackage[T1]{fontenc}

\usepackage[utf8]{inputenc}

\usepackage{microtype}

\usepackage{inconsolata}

\usepackage{pgf}
\usepackage{caption}
\usepackage{subcaption}
\usepackage{todonotes}

\newcommand{\model}{\mbox{\textsc{Ditto}}}
\newcommand{\smi}{\mbox{\textsc{Smi}}}

\newcommand{\figref}[1]{Fig.~\ref{#1}}
\newcommand{\tabref}[1]{Tab.~\ref{#1}}
\newcommand{\secref}[1]{Sec.~\ref{#1}}
\newcommand{\AlgRef}[1]{Algorithm~\ref{#1}}

\title{DITTO: Data-efficient and Fair Targeted Subset Selection \\ for ASR Accent Adaptation}

\author{Suraj Kothawade$^{1}$\thanks{\enspace Equal contribution.}, Anmol Mekala$^{2*}$, Chandra Sekhara D$^2$, Mayank Kothyari$^2$, \\ \textbf{Rishabh Iyer$^1$, Ganesh Ramakrishnan$^2$, Preethi Jyothi$^2$} \\
$^1$ The University of Texas at Dallas, Dallas, USA\\
$^2$ Indian Institute of Technology Bombay, Mumbai, India}

\begin{document}
\maketitle
\begin{abstract}
State-of-the-art Automatic Speech Recognition (ASR) systems are known to exhibit disparate performance on varying speech accents. To improve performance on a specific target accent, a commonly adopted solution is to finetune the ASR model using accent-specific labeled speech. However, acquiring large amounts of labeled speech for specific target accents is challenging. Choosing an informative subset of speech samples that are most representative of the target accents becomes important for effective ASR finetuning. To address this problem, we propose \model\ (\textbf{D}ata-efficient and fa\textbf{I}r \textbf{T}argeted subse\textbf{T} selecti\textbf{O}n) that uses Submodular Mutual Information (\smi) functions as acquisition functions to find the most informative set of utterances matching a target accent within a fixed budget. An important feature of \model\ is that it supports fair targeting for multiple accents, \ie\ it can automatically select representative data points from multiple accents when the ASR model needs to perform well on more than one accent. We show that \model\ is 3-5 times more label-efficient than other speech selection methods on the Indic-TTS and L2 datasets.
\end{abstract}

\section{Introduction}
State-of-the-art speech recognition systems have seen tremendous progress in the last few years, with end-to-end architectures becoming a default modeling choice. While end-to-end models yield impressive Word Error Rates (WERs) and work well for certain user populations~\cite{rao2017exploring,chiu2018state}, they severely underperform when confronted with out-of-domain test utterances in target accents that are unseen or rarely seen during training~\cite{feng2021quantifying, koenecke2020racial}. 

A common solution~\cite{shor2019personalizing,sim2019investigation} to address such mismatched settings is to adapt a well-trained, speaker-independent ASR model with a small amount of accent-specific target data to adapt models to the target setting. While these works propose different fine-tuning schedules that would be most beneficial given the limited amount of target data, the question of which utterances should be chosen in order to be transcribed and further used for fine-tuning has received far less attention. This is extremely important, since procuring and labeling accent-specific data is challenging and expensive. Awasthi {\em et. al.}~\cite{awasthi2021error} present a method to select sentences within a fixed budget that are most likely to induce ASR errors to record accented audio on, resulting in higher-quality personalized ASR models for target accents compared to random selection. However, they assume access to a small seed set of labeled utterances from the target speaker. We address a more realistic setting wherein we have access only to a limited number of \emph{unlabeled} utterances from the target domain, and without access to accented speakers to read out the selected texts.

\subsection{Our Contributions}

In this work, we propose \model\, a data-efficient and fair targeted subset selection approach that makes use of a suite of submodular mutual information (\smi) functions (originally defined in~\cite{iyer2021submodular}). For a specific target accent, we are given access to a small number (\textit{20} in our experiments) of unlabeled speech utterances, called the target (or query) set. We aim at identifying the most informative subset of speech utterances from a large unlabeled pool of diverse accents that best matches the target set. We procure the best matching subset by maximizing an \smi\ function instantiated using pairwise similarities between speech representations. We find \model\ to be an effective targeted subset selection technique for adapting ASR models in accents at multiple granularities - within Indian accents and accents around the world. \model\ uses a limited transcription budget, {\em i.e.}, just around 20-35\% of that of random. Furthermore, we show that \model\ can fairly select subsets that can cover multiple target accents using a facility location based \smi\ function.

\section{Related Work}
A number of works have studied subset selection for speech recognition. \citet{wei2014submodular,wei2014unsupervised,wei2013using} use submodular function-based subset selection on generated transcripts to find a minimal set of ASR training data and \citet{wu2007data} use an entropy measure for the same. \citet{asami2015training} employ a joint Kullback-Leibler divergence-based subset selection on out-of-domain samples for ASR adaptation across acoustic characteristics such as speaker, noise and recording devices. 
Similarly, \citet{liu2015svitchboard} study subset selection to obtain low-vocabulary speech corpora for ASR, while~\citet{kirchhoff2014submodularity} use a submodular approach for data selection in machine translation. Many recent papers~\cite{yuan2019gradient, syed2017active} have studied uncertainty and gradient based approaches for active learning to reduce the transcription time for ASR models, while \citet{hamanaka2010speech} use a committee-based active learning method to select speech utterances. 

A number of approaches have studied adaptation to atypical speech patterns like accented and dysarthic speech, such as~\cite{shor2019personalizing} and \cite{tomanek2021device} which fine-tune a subset of layers using labeled data from targeted accents. \citet{sun2018domain} employ domain adversarial training to adapt across accents. \citet{awasthi2021error} tries addressing a problem that corresponds exactly to the reverse of our setting by trying to determine the sentences a model is most error-prone to, and recording utterances for them. While this can be effective for user-driven personalization, our method is suited to settings in which we have fixed speech utterances, and the only actionable item for us is to transcribe a subset of them. All these approaches need data specifically from the target domain to be labeled toward use for training/fine-tuning.

Finally, a number of recent works on data selection have leveraged the submodular mutual information functions used in this work for targeted subset selection. \citet{kaushal2020unified} employ the SMI functions for query focused and privacy-preserving summarization, while \citet{kothawade2022prism} utilize the SMI functions for improving the model performance on targeted slices. Recently, \citet{kothawade2021similar} proposed an active learning approach using the SMI functions for rare classes, redundancy, and OOD data.

\section{Submodular Mutual Information (SMI) Functions} \label{sec:smi}

\label{sec:SMI-optimise}
\textbf{Submodular Functions:} We let $\Ucal$ denote the \emph{ground-set} of $n$ data points $\Ucal = \{1, 2, 3,...,n \}$ and a set function $f:
 2^{\Ucal} \xrightarrow{} \Re$. 
 The function $f$ is submodular~\cite{fujishige2005submodular} if it satisfies the diminishing marginal returns, namely $f(j | \Xcal) = f(\Xcal \cup j) - f(\Xcal) \geq f(j | \Ycal)$ for all $\Xcal \subseteq \Ycal \subseteq \Ucal, j \notin \Ycal$.
Submodularity ensures that a greedy algorithm achieves bounded approximation factor when maximized~\cite{nemhauser1978analysis}.

\noindent
\textbf{Submodular Mutual Information (SMI):} Given a set of items $\Scal, \Tcal \subseteq \Ucal$, the submodular mutual information (SMI)~\cite{levin2020online, iyer2021submodular} is defined as $I_f(\Scal; \Tcal) = f(\Scal) + f(\Tcal) - f(\Scal \cup \Tcal)$. Intuitively, this function measures the similarity between $\Tcal$ and $\Scal$ and we refer to $\Tcal$ as the targeted set. 
In the setting considered in this paper, the set $\Tcal$ (target set, also called query set) consists of a small set of unlabeled utterances from an accent, and $\Ucal$ is a large unlabeled set of utterances from multiple accents. To find an optimal subset given a target set $\Tcal$, we can define $g_{\Tcal}(\Scal) = I_f(\Scal; \Tcal)$, $\Scal \subseteq \Ucal$ and maximize the same. Using a greedy algorithm, these submodular functions can be efficiently optimized within an approximation factor (1-1/$e$)
of the global maximum.

\begin{figure*}[ht!]
\centering
\includegraphics[width = 0.85\textwidth]{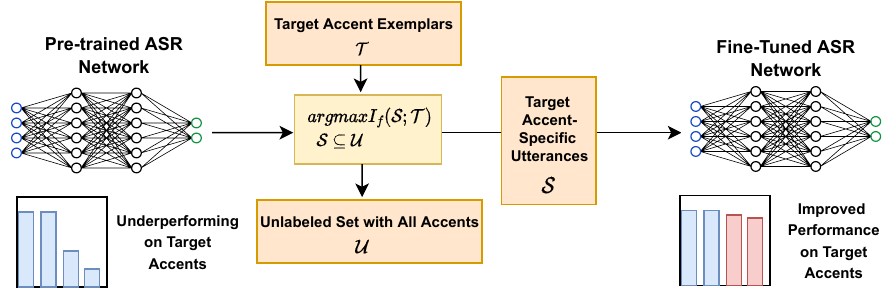}
\caption{ASR Accent Adaptation using \model.}
\label{fig:Ditto}
\end{figure*}

\subsection{SMI functions used in \model}
\label{sec:SMI-fun}
We use the SMI functions recently introduced in~\cite{iyer2021submodular} and their extensions introduced in \cite{kaushal2020unified, kothawade2022prism}. 
For any two data points $i \in \Ucal$ and $j \in \Tcal$, let $s_{ij}$ denote the similarity between them. \\

\noindent\textbf{Graph Cut MI:} The submodular mutual information (SMI) instantiation of graph-cut (GCMI) is defined as~\cite{kothawade2022prism, iyer2021submodular}:
\begin{align} \label{eq:GCMI}
I_f(\Scal;\Tcal)=2\sum_{i \in \Scal} \sum_{j \in \Tcal} s_{ij}
\end{align}
Since maximizing GCMI maximizes the joint pairwise sum with the query set, it will lead to a summary similar to the query set $Q$. GCMI models only query-relevance and does not select based on diversity~\cite{kothawade2022prism}. \\

\noindent\textbf{Facility Location MI:} 
The Facility Location Mutual Information (FLMI) function~\cite{kothawade2022prism} takes the expression:
\begin{align} \label{eq:FLMI}
I_f(\Scal;\Tcal)=\sum_{i \in \Tcal} \max_{j \in \Scal} s_{ij} + \sum_{i \in \Scal} \max_{j \in \Tcal} s_{ij}
\end{align}
FLMI jointly models representation and query-relevance. It measures a bidirectional similarity between representation of data points that are the most relevant to the query set, and vice versa. \\

\section{\model: Our Data-efficient and Fair Targeted Subset Selection Method} \label{sec:our_method}

In this section, we discuss \model\, our data-efficient and fair targeted subset selection method for ASR accent adaptation. We show that \model\ can select fair and target-relevant subsets, which is critical for fine-tuning ASR models on one or more accents. The main idea of our method is to instantiate a submodular mutual information (SMI) function using appropriate similarity kernels in order to jointly optimize it for targeting and fairness. We summarize our method in \AlgRef{algo:TSSforSpeech} and illustrate it in \figref{fig:Ditto}.

\begin{algorithm}
\begin{algorithmic}[1]
\REQUIRE Target $\mathcal{T}$, Budget $B$, \smi\ function type $f$, large unlabeled set $\Ucal$, Accent representation model $\Fcal$, ASR model $M$ with parameters $\theta$\\
\STATE $\Ecal_\Tcal \leftarrow $ $\Fcal(\Tcal$) \textcolor{blue}{\{$\Ecal_\Tcal \in \mathbb{R}^{|\Tcal| \times D}$\}}
\STATE $\Ecal_\Ucal \leftarrow $ $\Fcal(\Ucal$) \textcolor{blue}{\{$\Ecal_\Ucal \in \mathbb{R}^{|\Ucal| \times D}$\}}
\STATE $\Xcal \leftarrow $ \textsc{Similarity}($\Ecal_\Tcal, \Ecal_\Ucal$) \textcolor{blue}{\{$\Xcal \in \mathbb{R}^{|\Tcal| \times |\Ucal|}$\}}
\STATE Define an SMI function $g_{\Tcal}(\Scal) = I_{f}(\mathcal{S};\mathcal{T})$ using $\Xcal$
\STATE $\mathcal{S} \gets \argmax\limits_{\mathcal{S}\subseteq\mathcal{U},c(\mathcal{S}) \leq B} g_{\Tcal}(\Scal)$ \\ \textcolor{blue}{\{\small Greedy maximization of $g$ to select a subset $\Scal$\}}\\
\STATE $D \gets $ Transcribe utterances in $\mathcal{S}$\\
\STATE $\hat{\theta} \leftarrow $ Fine-tune ASR model $M$ on $D$\\
\STATE \textbf{Return} Fine-tuned model $\Mcal$ with updated parameters $\hat{\theta}$.
\end{algorithmic}
\caption{\model\ for Adapting ASR model $M$}
\label{algo:TSSforSpeech}
\end{algorithm}

Concretely, we are provided a few unlabeled utterances from the accent (a target set $\mathcal{T}$) which we would like the ASR model $M$ to be adapted to. The goal is to select the most informative subset $\Scal$ \textit{with respect to} a target $\Tcal$ from a large corpus $\mathcal{U}$ of unlabeled data, called the \textit{ground set}. We are given a budget constraint, which is a constraint on the total time of the selected utterances. This corresponds to the transcription budget, since the selected utterances need to be later transcribed by a human.

We begin with extracting accent feature representations of the unlabeled set $\Ucal$ and the target set $\Tcal$; we subsequently discuss the feature representation in \secref{sec:exptsetup}. Next, we compute a similarity matrix $\Xcal$, which is an RBF kernel containing pairwise similarities $\Xcal_{ij}$ between all data points in $i \in \Tcal$ and $j \in \Ucal$. We use $\Xcal$ to instantiate one of the \smi\ functions $I_f(\Scal; \Tcal)$ discussed in \secref{sec:smi}. Specifically, we optimize $g_{\Tcal}(\Scal) = I_f(\Scal; \Tcal)$ for $\Scal \subseteq \Ucal$ subject to the cardinality constraint $c(\Scal) \leq B$, where $c$ corresponds to the duration (in seconds) of the specific utterance and $B$ is the time budget. We use the greedy algorithm~\cite{mirzasoleiman2015lazier, nemhauser1978analysis, lin2010multi} with memoization~\cite{iyer2019memoization} and with a knapsack constraint on the optimization. Specifically, given the current set $\Scal$, we select the item $i = \argmax_{j \in \Ucal \backslash \Scal} g_{\Tcal}(j | \Scal)$, with the stopping criterion as $c(\Scal) \leq B$. Once, we obtain the set $\Scal$ as the solution of this optimization problem, we obtain $\Scal$'s transcriptions from a human, and fine-tune the ASR model using $\Scal$ and its labels.\\

\noindent \textbf{Scalability of \model:} The selection time of \model\ is dominated by the instantiation and maximization of the \smi\ function. Since all \smi\ functions used in this work are graph based, they require the computation of a similarity kernel. Hence, the main components that contribute towards the time complexity of \model\ are the similarity kernel computation and the greedy maximization. The FLMI and GCMI functions require a $t \times u$ similarity matrix, where $t$ =  $|T|$ is the number of points in the target set and $u$ = $|U|$ is the number of points in the unlabeled ground set. This leads to a $O(tu)$ complexity for computing the kernel. Given a selection budget of $B$, the time complexity of the greedy maximization for FLMI and GCMI is $O(tuB)$, which is linear in budget and ground set sizes.

\section{Experimental Setup}
\label{sec:exptsetup}

\subsection{Datasets} \label{sec:datasets}

We experiment with adapting ASR models on two public datasets, {\em viz.}, \textit{IndicTTS} and \textit{L2-Arctic}, containing English speech in various non-native accents. IndicTTS \cite{vignesh2016significance} consists of 35K utterances from 8 Indian speakers, each with a different accent depending on their native language: Gujarati (GUJ) 9.2\% of samples, Kannada (KAN) 9.4\%, Tamil (TAM) 15\%, Malayalam (MAL) 10.1\%, Hindi (HIN) 10.1\%, Rajasthani (RAJ) 9.5\%, Assamese (ASM) 16.7\% and Manipuri (MAN) 20.1\%. L2-Arctic~\cite{zhao2018l2} has 18K samples of accented English speech from 24 speakers spanning six non-native accents: Hindi (HIN), Vietnamese (VTN), Chinese (CHN), Korean (KOR), Arabic (ARB) and Spanish (ESP). The distribution among the accents is uniform for this datasets, with all represented equally.

\noindent \textbf{Feature representation:} Each utterance is represented as a 39-dimensional vector of MFCC coefficients averaged over the duration of the utterance.


\subsection{ASR Model Description and Fine-tuning Details}

Following~\cite{awasthi2021error}, our pre-trained model is based on the QuartzNet-15x5~\cite{kriman2020quartznet} architecture. It is trained on LibriSpeech~\cite{panayotov2015librispeech} for 400 epochs using the CTC-loss~\cite{graves2006connectionist} and yields a Word Error Rate (WER) of 3.90 on the test-clean split of LibriSpeech. The QuartzNet-15x5 architecture is fully convolutional with residual connections. This model is fine-tuned with our selected targeted subsets $\Scal$ of accented speech to minimize CTC loss using the NovoGrad optimizer~\cite{ginsburg2019stochastic} for 100 epochs with a batch size of 16, a linearly decaying learning rate of 10$^{-5}$ and early stopping based on the dev set. In all our experiments, we report results averaged over three runs using three different seeds and report error bars in all plots. We used an NVIDIA GTX 1080 Ti GPU for all runs.

\section{Experimental Procedure and Results}
 We use a transcription budget of 20.5 minutes for single-accent targeting and 41 minutes when an accent pair is targeted. The average uttterance durations are 4.92s in IndicTTS and 3.6s in L2-Arctic, thus these budgets come out to 250 and 500 samples on IndicTTS and 340 and 780 samples on L2-Arctic respectively.

  In our proposed method, we use the approach outlined in \AlgRef{algo:TSSforSpeech}, with the SMI function $\Ical_f$ set as one of the FLMI or GCMI functions. We consider them since they are computationally efficient (see \secref{sec:our_method}), and model different characteristics in their selections. As discussed in \secref{sec:SMI-fun}, GCMI models \textit{only} query-relevance, while FLMI models \textit{both} query-relevance and diversity. As we shall see in \secref{sec:single-acc} and \secref{sec:two-acc}, GCMI is an apt choice for targeting in some scenarios, whereas FLMI outperforms all methods when fairness and diversity are to be jointly modeled with targeting.

\input{tables/indic/indic-1.tex}
\input{tables/l2/l2-1.tex}

 \subsection{Baseline selection approaches}
We compare the performance of \model\ with a wide-array of standard selection techniques designed for subset selection in speech. In our results, we track the improvements in WER over the performance of the pretrained ASR model (without any finetuning), denoted as ``Pre". For all single/multiple accent experiments, we compare \model\ with the following selection approaches: \\

    \noindent \textbf{Random}: Selecting utterances randomly from the different accents of the ground set. The selection distributions will roughly match the ground set.
    
    \noindent \textbf{phone-diverse (``PhoneDiv")}: Following an approach from \cite{awasthi2021error}, we select a phonetically rich set from the set of generated transcripts of our baseline ASR model on the utterances. We define a diminishing returns submodular set-scoring function that penalizes over-represented phones. Similar to optimizing SMI functions, greedy optimization~\cite{mirzasoleiman2015lazier} of this function gives us our selected subset. The function score$(\Scal)$ is defined as follows, with $\Scal$ denoting a set of sentences, $\Pcal$ denoting the set of phones, $\tau$ a hyperparameter for the penalization (set to 500 in our experiments), $n_{\pi}(\Scal)$ denoting the number of utterances of phone $\pi$ in $\Scal$:
    \begin{align} \label{eq:phone-score}
        \textnormal{score}(\Scal)=\sum_{\pi}(1-\textnormal{exp}(-n_{\pi}(\Scal)/\tau))
    \end{align}
    
    \noindent \textbf{Maximum ASR entropy (``Entropy")}: This method is from \citet{riccardi2005active} and focuses on selecting utterances that the baseline ASR model is 
    most uncertain about. We score each utterance by finding the entropy across frame-level phone predictions and averaging across frames. 
    We then pick the highest scoring utterances within our budget duration.  The score of an utterance $\Ccal$ is defined as follows, with $\Fcal$ denoting the frames of the utterance $\Ccal$, $p_f(\pi)$ denoting the ASR model's softmax value on phone-$\pi$ of frame-$f$. 
    
    \begin{align} \label{eq:entropy-score}
        \textnormal{score}(\Ccal)=\frac{1}{|\Fcal|}\sum_{\textnormal{frame} f \in \Fcal}\sum_{\pi}{-p_f(\pi) log(p_f(\pi))}
    \end{align}
     We also use two submodular functions that are well-known for subset selection tasks. Namely, Facility Location and Log Determinant functions.\\
\noindent \textbf{Facility Location (``FL")}: The facility location function is known to select a representative subset and has been extensively used for speech data subset selection tasks \cite{wei2014submodular,wei2013using}. Using the same notation as in \secref{sec:our_method}, where $\Scal$ denotes the subset of utterances to be selected from the unlabeled set $\Ucal$, the FL function is defined as:

\begin{equation} \label{eq:FL}
    f(\Scal) = \sum_{i \in \Ucal} \max_{j \in \Scal} \Xcal_{ij}
\end{equation}

\noindent \textbf{Log Determinant (``LogDet")}: The log determinant function models diversity and is crucial for
determinantal point processes (DPPs). The LogDet function is defined as follows:

\begin{equation} \label{eq:lodget}
    f(\Scal) = \mbox{Log Det}(\Xcal_\Scal)
\end{equation}

\noindent where, $\mbox{Det}(.)$ is the determinant, and $\Xcal_\Scal$ denotes the rows and columns of the similarity matrix instantiated with elements in $\Scal$.

For fair evaluation, the FL and LogDet functions are optimized using the same greedy strategy \cite{mirzasoleiman2015lazier} as the SMI functions used in \model. Note that FL and LogDet functions are computationally expensive since they require the computation of a $O(n^2)$ similarity matrix, as opposed to the SMI functions, which can be optimized in linear time.

\begin{figure}[tp]
\begin{center}
    \includegraphics[width = 8cm, height=0.80cm]{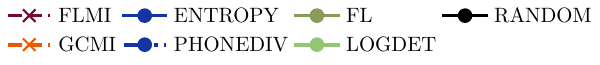}
    \begin{subfigure}{0.45\textwidth}
        \centering
        \includegraphics[width=\textwidth]{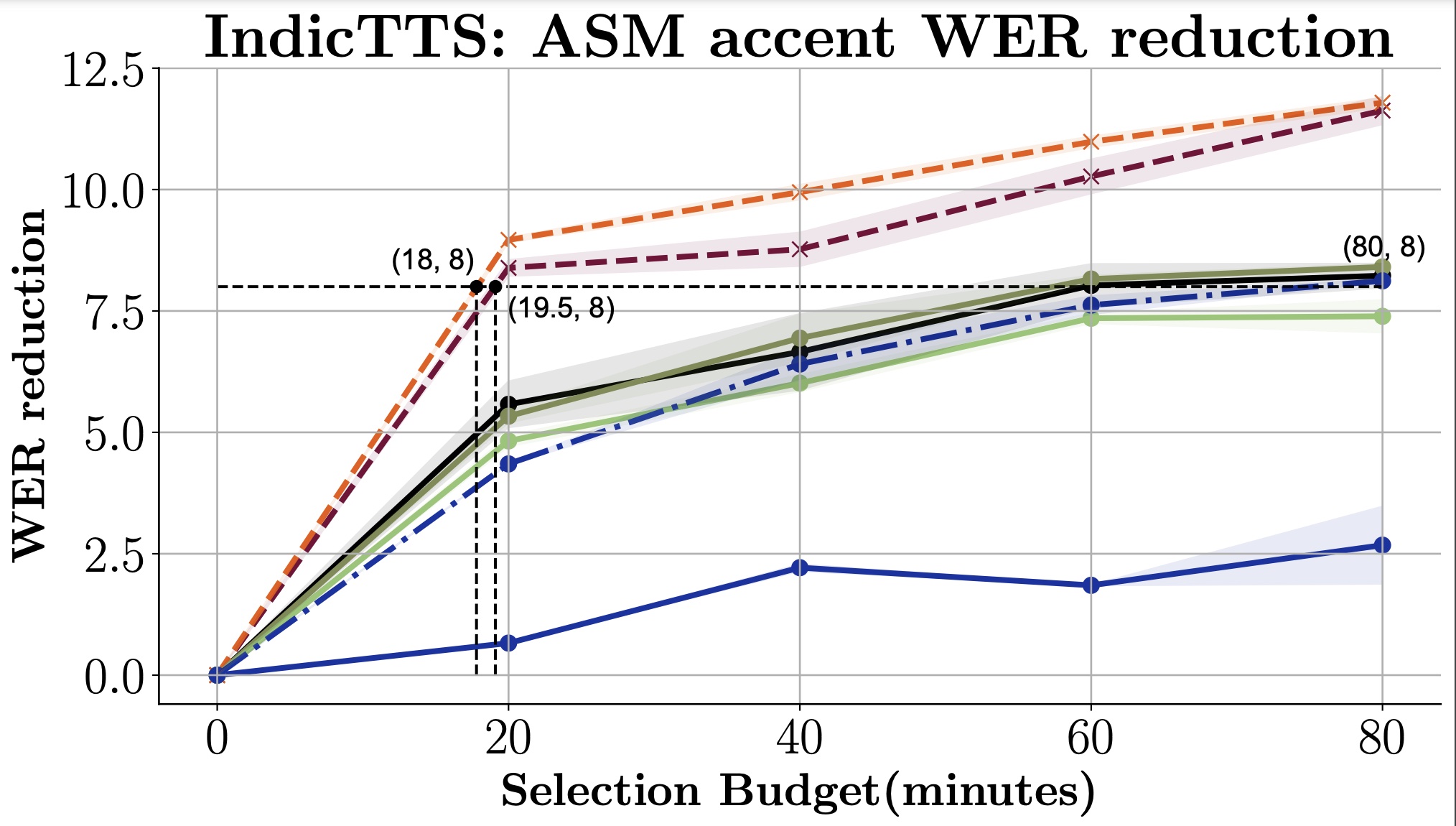}
    \end{subfigure}
    \begin{subfigure}{0.4\textwidth}
        \centering
        \includegraphics[width=\textwidth]{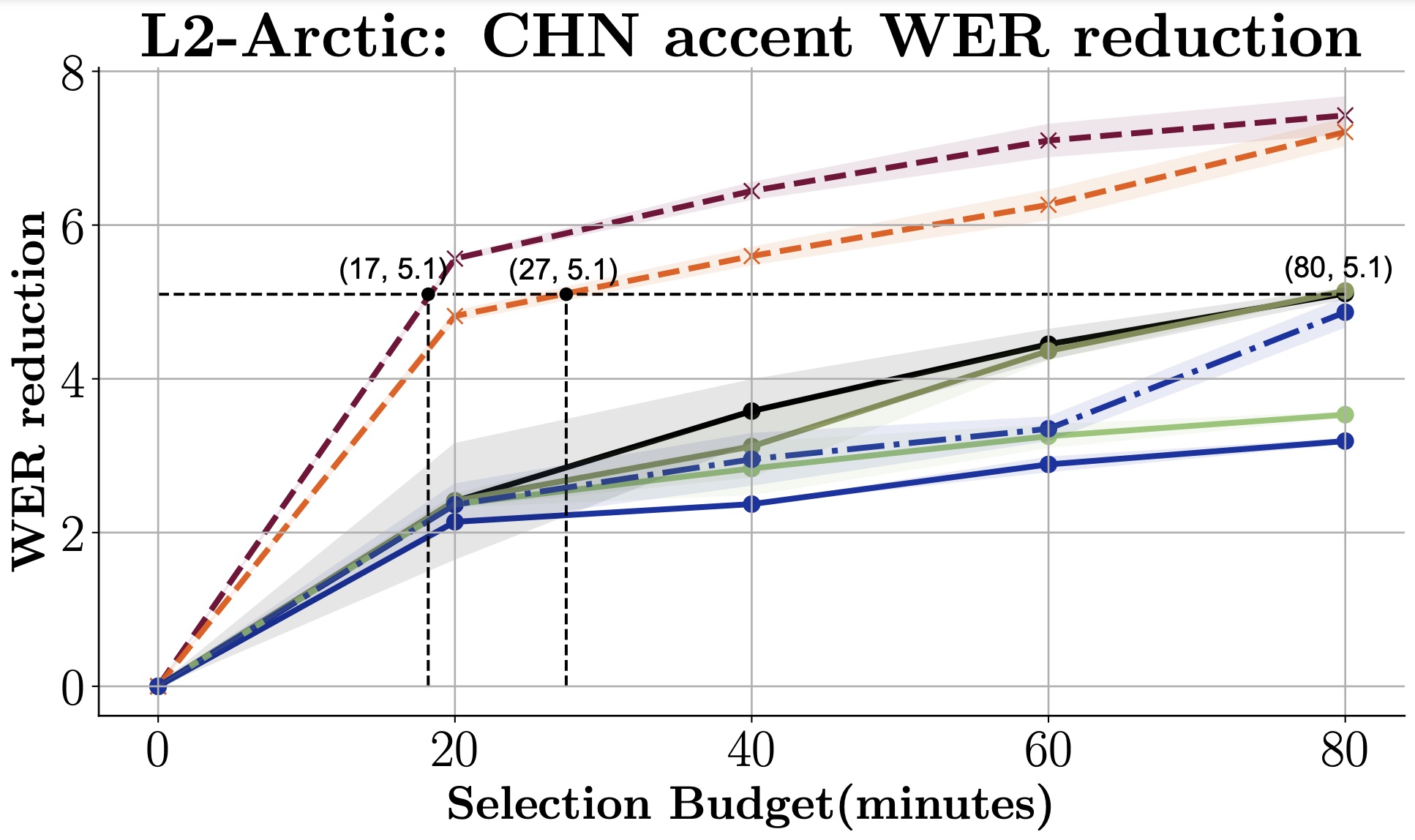}
    \end{subfigure}
\end{center}
\caption{WER reductions across a range of budgets, targeting Assamese (from IndicTTS) and Chinese (from L2-Arctic) accents.}
\label{fig:singleAcc-budget}
\end{figure}

\subsection{Targeted Subset Selection for Single-accents}
\label{sec:single-acc}

In this section, we analyze the performance of \model\ for procuring a subset that is targeted for a single accent, followed by fine-tuning an ASR model using the selected targeted subset. For evaluation, we study the \textit{Word Error Rate} (WER) of the targeted accent by evaluating the fine-tuned model on a held-out test set containing utterances of the targeted accent. Along with the WER, we also report a \textit{Targeted percentage} ($\mathbf{T}$\%) that denotes the ratio of utterances selected from the targeted accent given the total budget. 

We conduct extensive experiments on IndicTTS \cite{vignesh2016significance} and L2-Arctic \cite{zhao2018l2} datasets (see \secref{sec:datasets} for details) by targeting all accents in both datasets, one accent at a time. For each accent (around 4.3K samples in IndicTTS and 3K samples in L2-Arctic), we create data splits by partitioning $70\%$ of the data for the unlabeled set ($\Ucal$) and a small target set $\Tcal$ of size 20. Of the remaining 30\%, we create a test set from 27\% and use 50 samples from the 3\% as the finetuning dev set. In L2 Arctic, which has equal samples from each speaker of an accent, we ensure an equal split across the speakers in our accent-specific query, test and dev sets.

\input{tables/indic/Combined.tex}
\begin{figure*}[t!]
    \centering
      \includegraphics[width = 16cm, height=0.5cm]{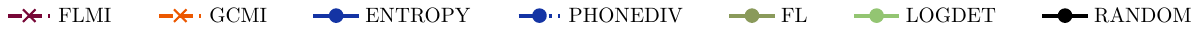}
    \begin{subfigure}{0.325\textwidth}
        \includegraphics[width=\textwidth]{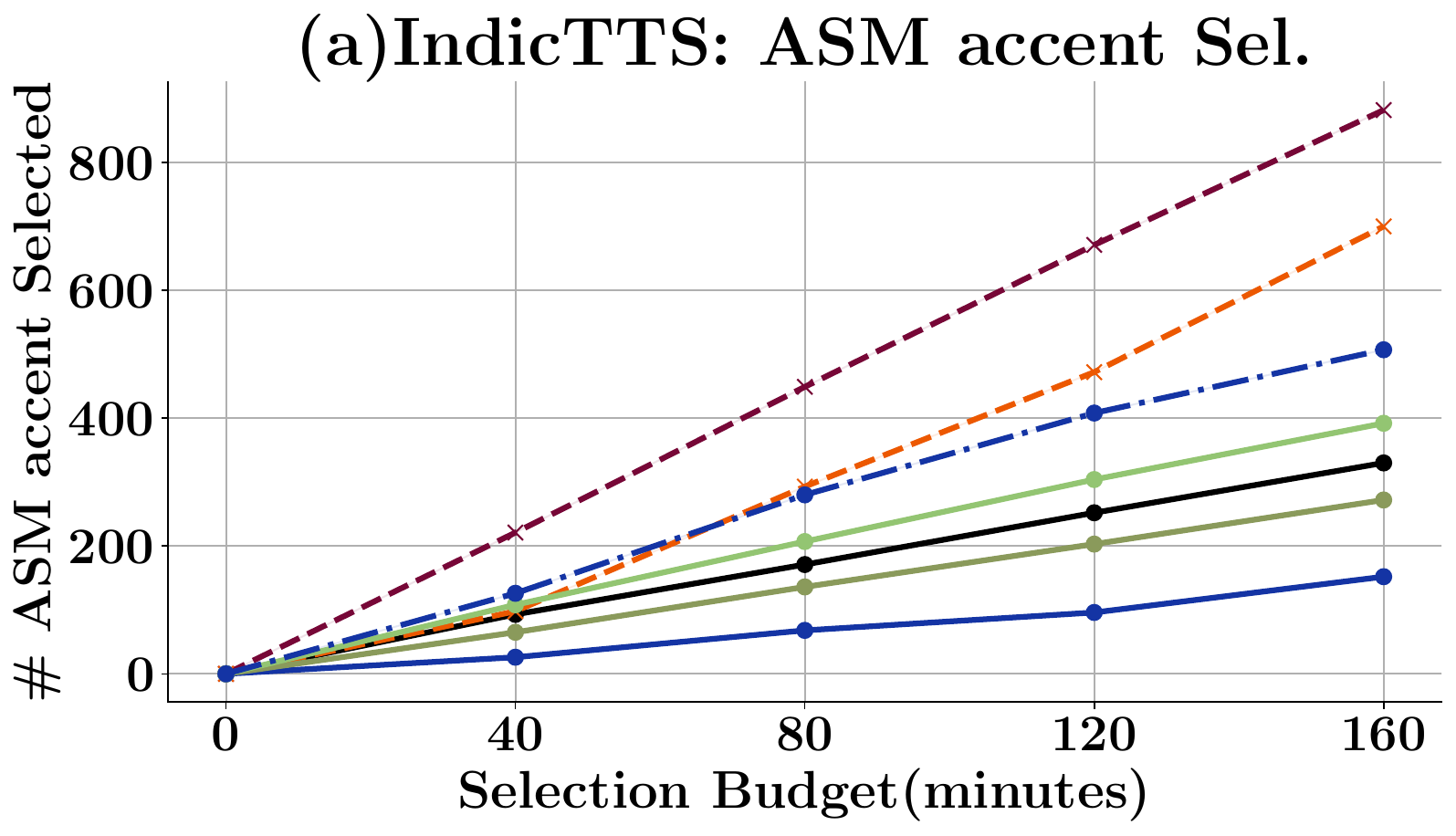}
    \end{subfigure}
    \begin{subfigure}{0.325\textwidth}
        \includegraphics[width=\textwidth]{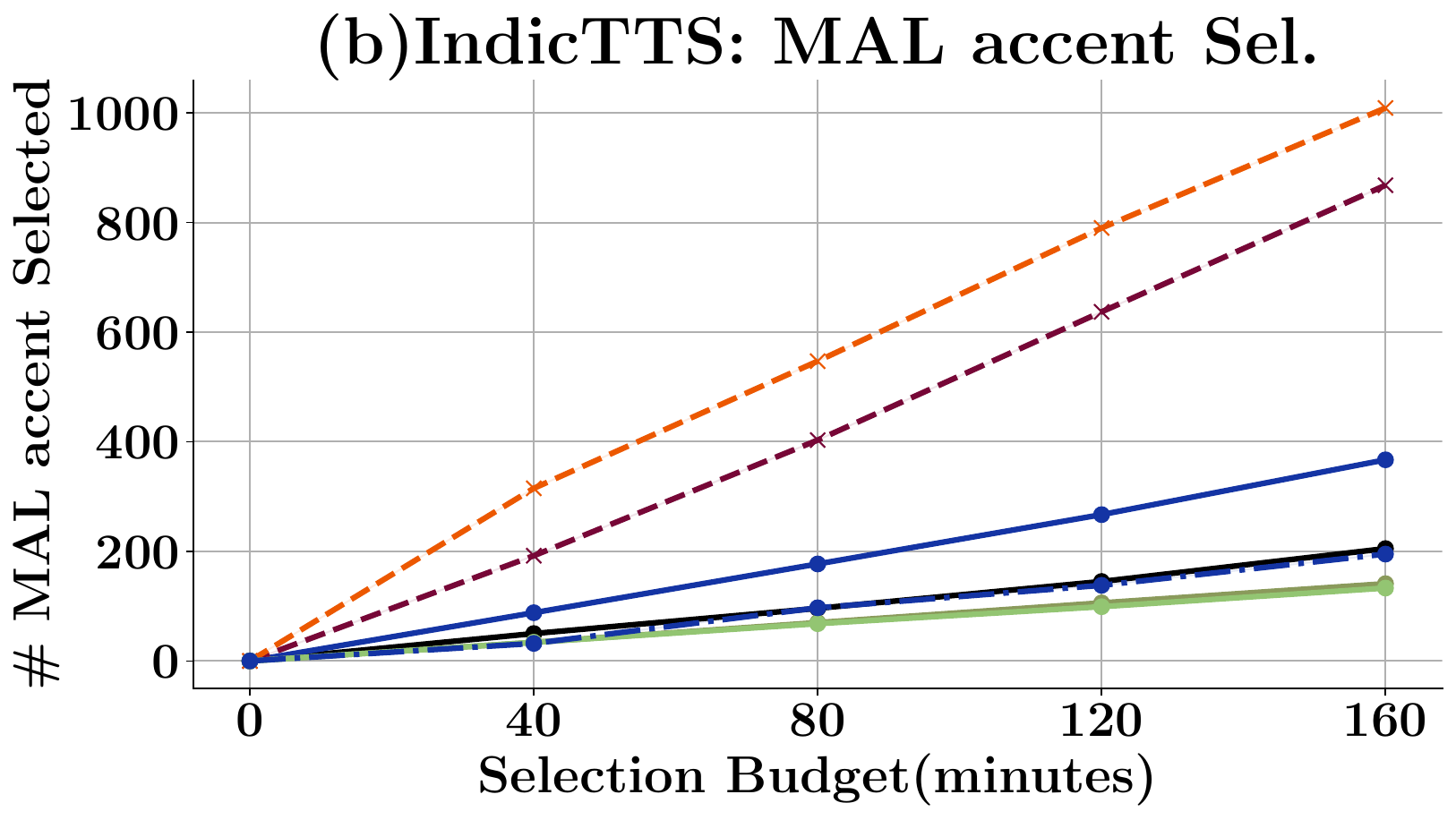}
    \end{subfigure}
    \begin{subfigure}{0.325\textwidth}
        \includegraphics[width=\textwidth]{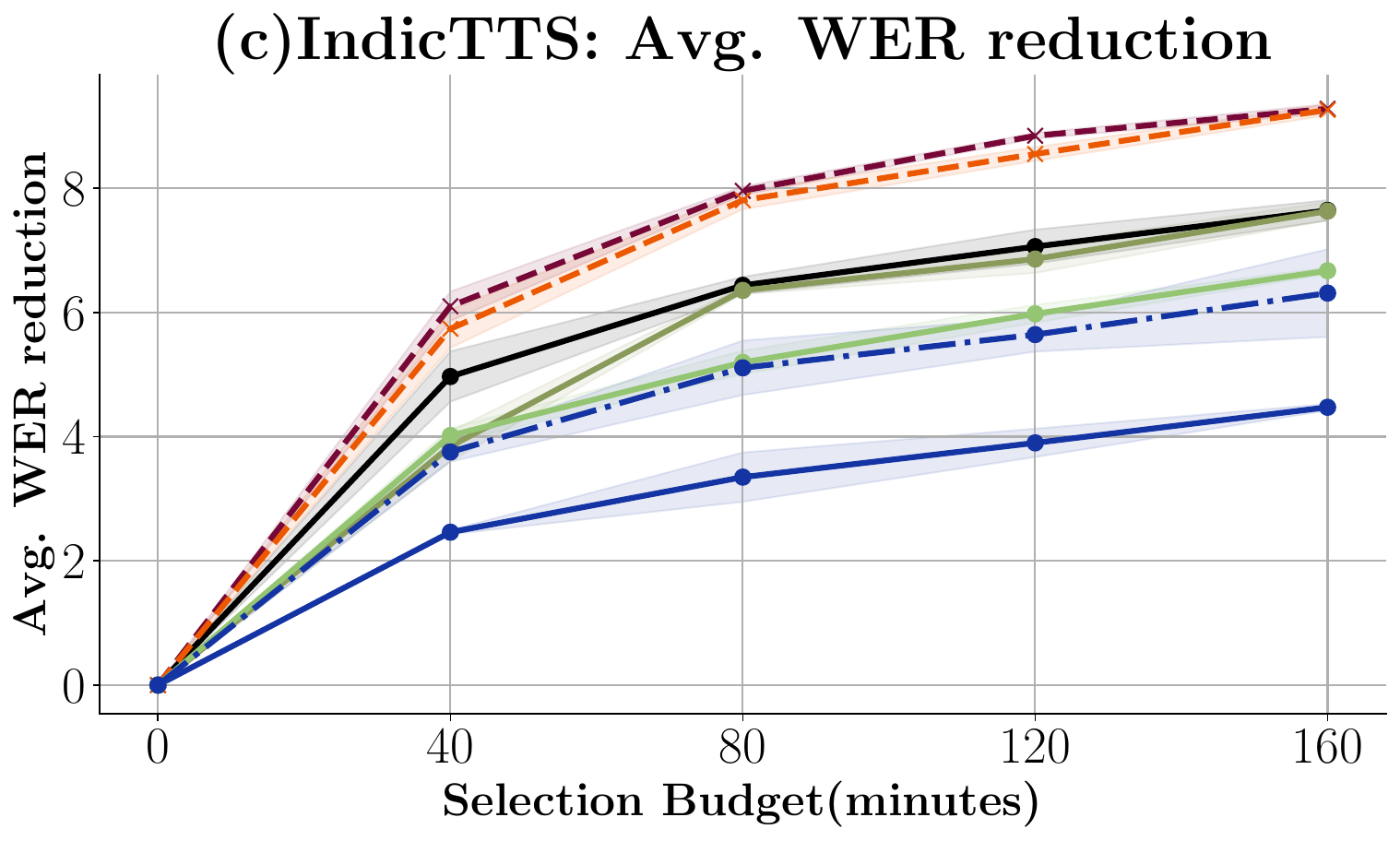}
    \end{subfigure}
  \caption{Variation in selections and average WER across a range of budgets, targeting ASM and MAL together.}
  \label{fig:indic-multi-plot}
\end{figure*}

We present the targeted subset selection and fine-tuning results for the IndicTTS dataset in \tabref{tab:indic-1} and for L2-Arctic in \tabref{tab:l2-1}. We observe that the SMI functions, GCMI and FLMI outperform other methods in terms of WER for all target accents. This is due to the fact that the SMI functions are able to identify utterances from the target accent almost to perfection in most cases. Interestingly, GCMI performs better than FLMI on IndicTTS in 6 out of 8 accents due to its high predilection towards query-relevance. On the other hand, GCMI performs worse than FLMI (although better than other methods) in terms of WER on the L2-Arctic dataset. This is because FLMI is significantly better in terms of targeting in comparison to GCMI and other methods. Note that IndicTTS is simpler since it contains data from only one speaker per accent, whereas L2-Arctic has comparatively more complex acoustics as it contains data from multiple speakers for each accent. We believe that the representation modeling capability and bidirectional similarity of FLMI allows for a better targeting ability on complex datasets like L2-Arctic. On the other hand, for datasets with lower acoustic complexity like IndicTTS: GCMI, which performs query-relevance, works well. We also present the variation in WER improvements across a range of budgets in \figref{fig:singleAcc-budget} for accents picked from each dataset. The horizontal lines marked indicate the how much budget each method needed for the same WER gain. For Assamese (ASM) we see that Random needs 80 minutes to improve by 8, while FLMI and GCMI do it in 18 and 19.5 mins respectively. For Chinese (CHN) we observe: for 5.1 gain, Random needs 80 minutes, while FLMI and GCMI need only 17 and 27 minutes respectively. The SMI functions are thus 3-5 times as label efficient than random.

\input{tables/l2/Combined.tex}

\begin{figure*}[ht!]
    \centering
     \includegraphics[width = 16cm, height=0.5cm]{plots/ditto_legend.pdf}
    \begin{subfigure}{0.325\textwidth}
        \includegraphics[width=\textwidth]{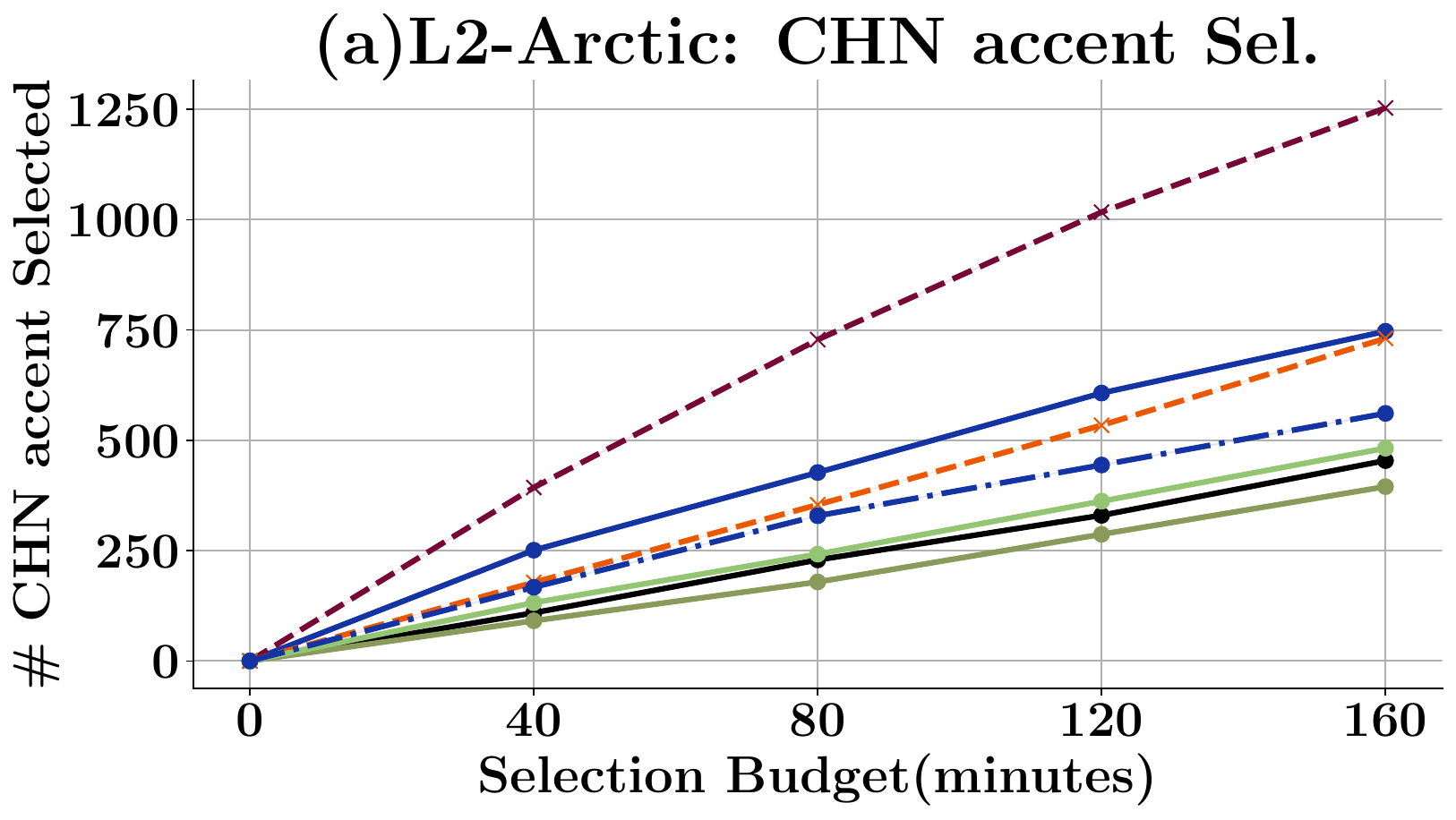}
    \end{subfigure}
    \begin{subfigure}{0.325\textwidth}
        \includegraphics[width=\textwidth]{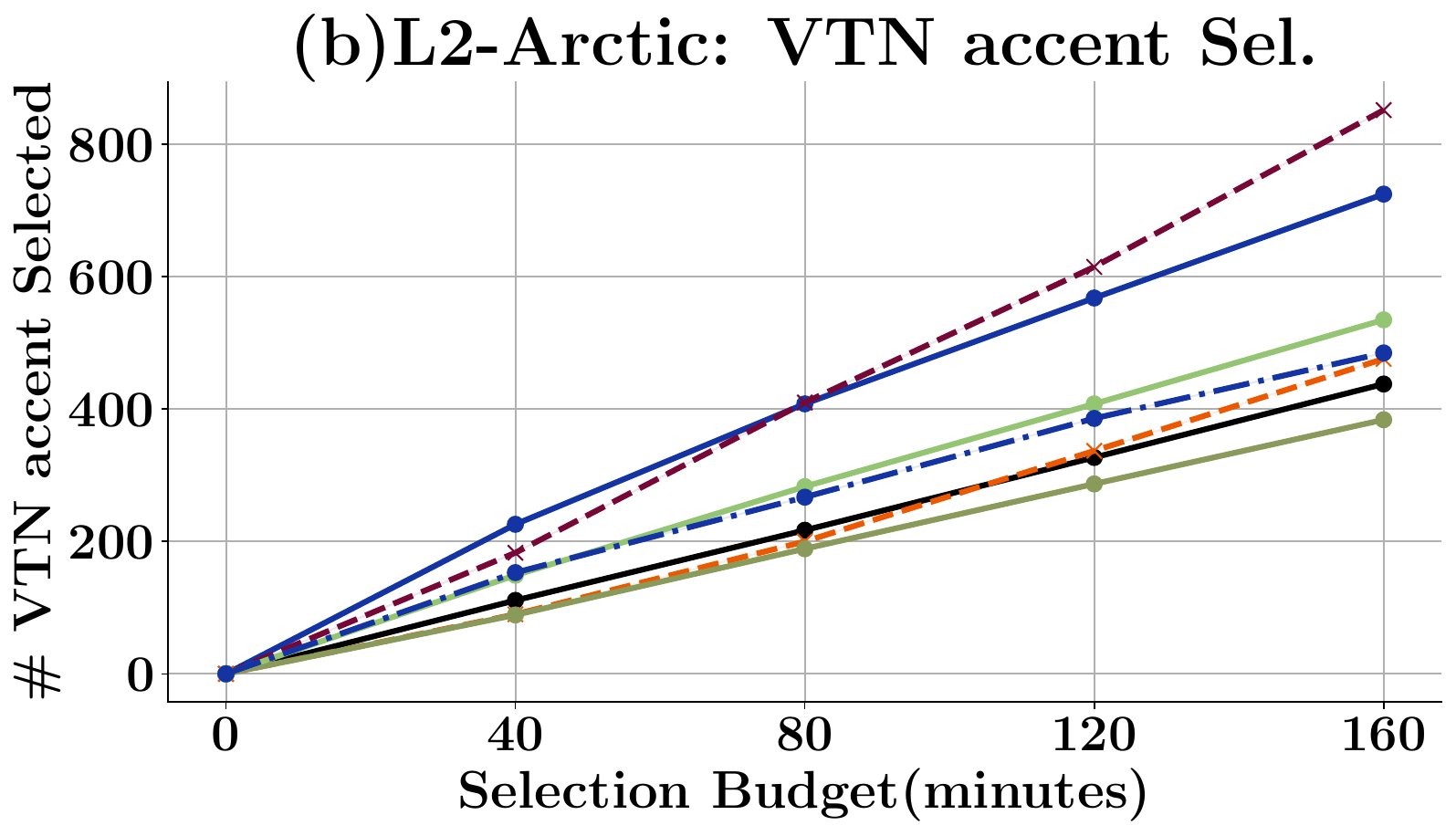}
    \end{subfigure}
    \begin{subfigure}{0.325\textwidth}
        \includegraphics[width=\textwidth]{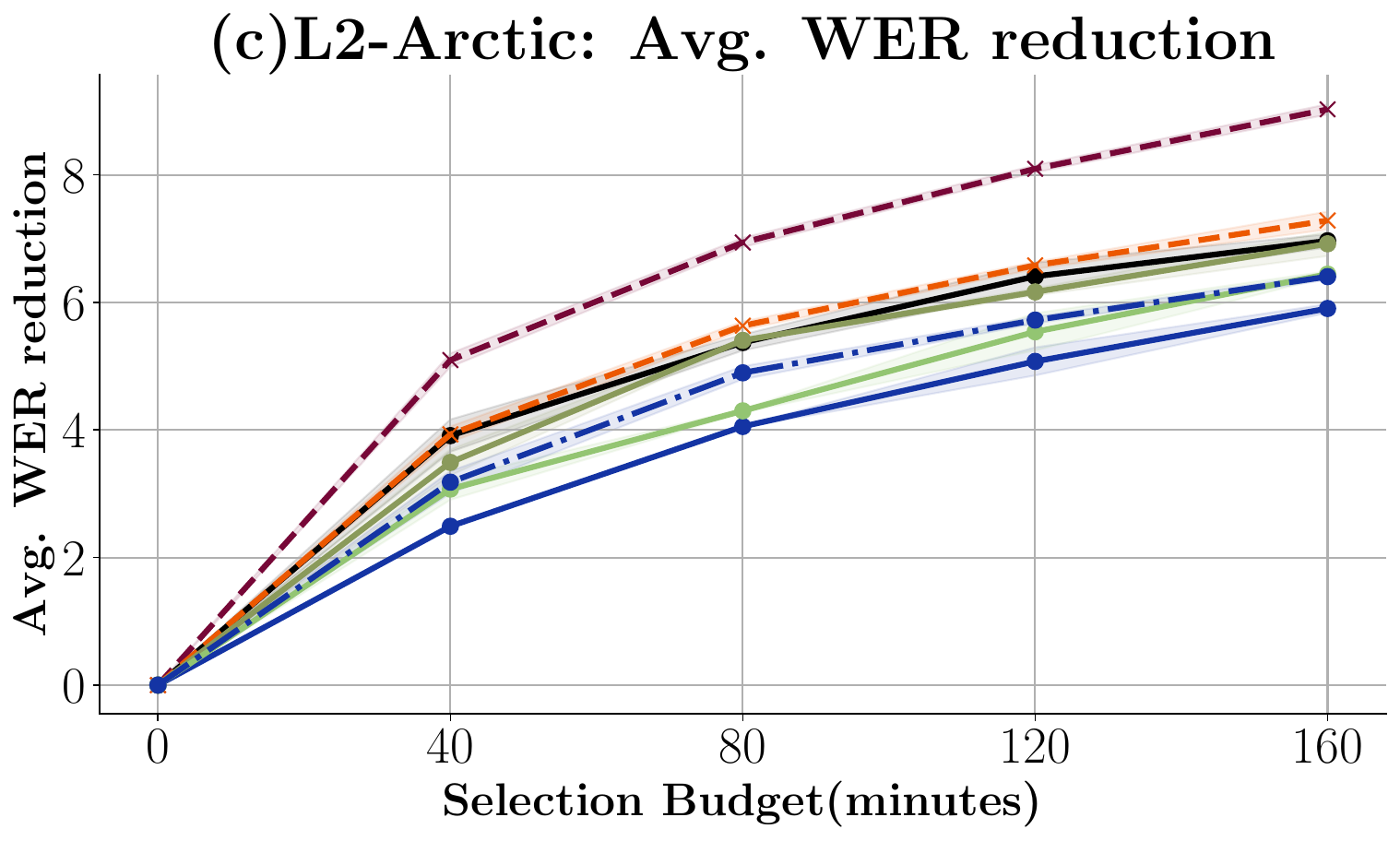}
    \end{subfigure}
  \caption{Variation in selections and average WER across a range of budgets, targeting CHN and VTN together.}
  \label{fig:l2-multi-plot}
\end{figure*}


\begin{figure*}[ht]
\centering
  \includegraphics[width=0.95\textwidth]{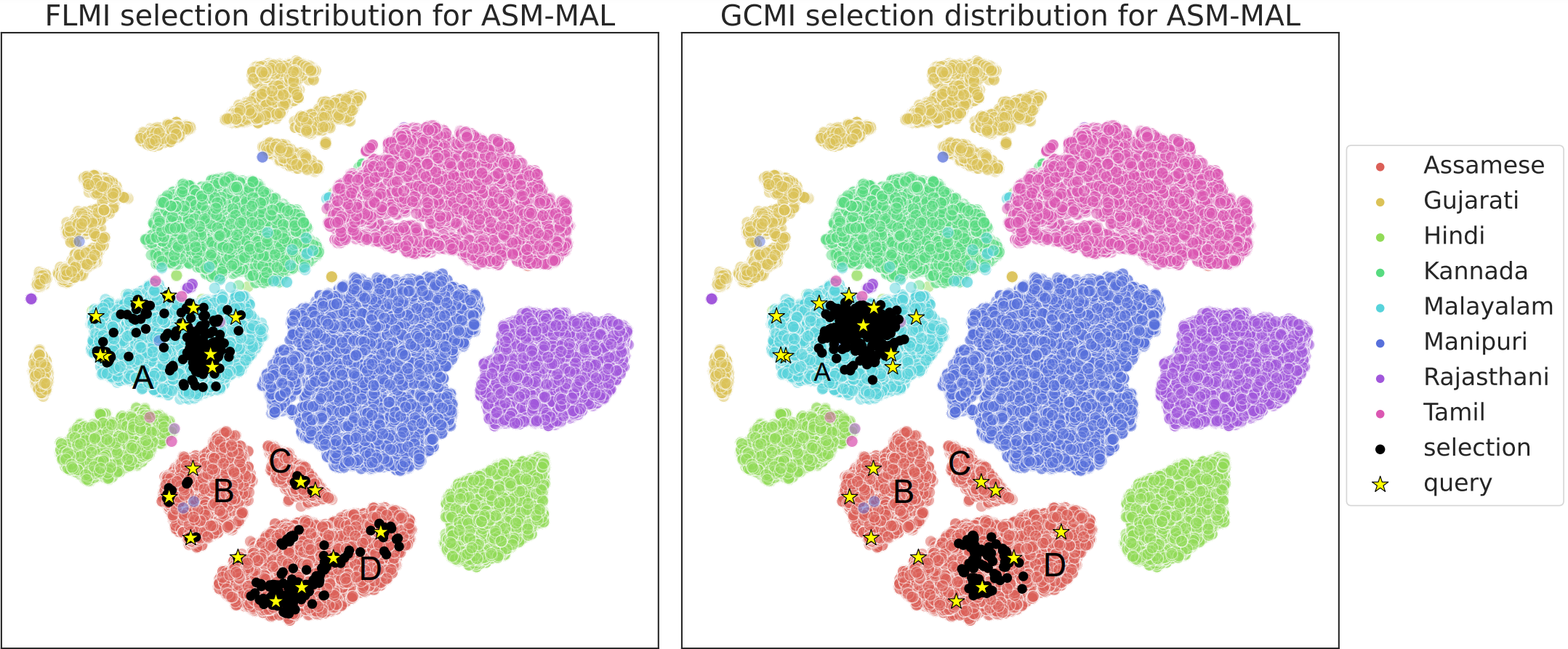}
  \caption{t-SNE visualising MFCC features of selections on the full IndicTTS dataset: comparing FLMI and GCMI. FLMI’s selections are representative of the query and spread over all clusters A, B, C and D. GCMI is biased towards the bigger cluster centers: it does not select from clusters B and C at all (thus selecting a lower Assamese fraction) and selects dense, unrepresentative clusters from the centers of A and D.}
  \label{fig:selection-tsne}
\end{figure*}

\subsection{Fair Targeted Subset Selection for Multiple-accents}
\label{sec:two-acc}

Another important setting of high practical value is that of adapting an ASR model for multiple targeted accents. In a real-world scenario, practitioners may want to improve the performance of the ASR model on accents that are under-performing. In another deployment scenario, one may need to fine-tune the ASR model on multiple accents in order to deploy it in a region where the population speaks in more than one accent. To tackle such scenarios, an ideal selection function would model fairness and select approximately equal number of utterances from each accent. To study this, we evaluate the performance of \model\ for targeting pairs of accents, followed by fine-tuning the ASR model on the selected targeted subset. For evaluation, we study the WER and average WER for both the targeted accents by evaluating the fine-tuned ASR model on separate individual held-out test sets containing utterances from each of the targeted accents. Similar to the single accent experiments, we also report the \textit{Targeted percentage} ($\mathbf{T\%}$) for each targeted accent. In addition, we report a \textit{Targeted Fairness} (\textbf{TF}) score for the accent-pair, which is computed as a product of the targeted percentages of both the targeted accents. We multiply the final score by 4 to obtain a TF score of 1 when the selected subset perfectly targets both accents, \ie\ it achieves 50\% targeted percentage for both the targeted accents.

 For our experiments, we consider pairs of the three worst-performing accents as target accent pairs from IndicTTS and L2-Arctic datasets. We present  results for three target accent pairs from IndicTTS in \tabref{tab:IndicTTS-2}: i) Assamese and Malayalam, ii) Malayalam and Rajasthani, and iii) Assamese and Rajasthani. We use data splits created in \secref{sec:single-acc}: ground and test sets remain the same, whereas query sets for the accent pairs here are made by taking 10 from each accent from the accent-specific query sets of \secref{sec:single-acc}.
 
 We observe that the SMI functions (GCMI and FLMI) outperform other methods in terms of the Avg. WER and the TF score. Interestingly, we also observe that GCMI often favors targeting a single accent: MAL when MAL and ASM are targeted, RAJ when RAJ and MAL are targeted and RAJ from RAJ and ASM. Due to this, GCMI obtains a lower TF score than FLMI. It is worth noting that FLMI achieves a TF score of as high as 1 (see \tabref{tab:IndicTTS-2}'s Asm-Mal section) due to its ability to jointly model representation. We find that GCMI tends to favor a particular accent $A$ due to higher pairwise similarity values $\Xcal_{A}$ between utterances belonging to accent $A$ in comparison to accent $B$. In \figref{fig:indic-multi-plot}, for the ASM-MAL accent pair, we illustrate the Avg. WER improvement and duration of utterances selected from both the accents across a wide range of budgets. Notably, we observe that FLMI continues to select fairly for both accents, while GCMI favors MAL accent over ASM. 

\input{tables/indic/Combined-Appendix.tex}

To compare the Targeted Fairness between FLMI and GCMI, we visualize a t-SNE plot of the IndicTTS dataset embedded using MFCC features in \figref{fig:selection-tsne}. As shown in the legend of \figref{fig:selection-tsne}, each color represents a particular accent and the selected data points are denoted in black. The query data points from ASM and MAL accents are shown by yellow stars. We observe that the data points selected by FLMI are representative of the query, as they are spread well across MAL (cluster A) and ASM (clusters B, C and D). On the other hand, data points selected by GCMI are mainly concentrated in the bigger cluster centers of MAL (cluster A) and ASM (cluster D), while completely missing clusters B and C. This is again a consequence of the fact that FLMI jointly models representation and query-relevance whereas GCMI focuses only on query-relevance.

We conduct a similar analysis for the L2-Arctic dataset. We present the results for pairs from three bottom performing target accents from L2-Arctic in \tabref{tab:l2-2}: i) Arabic and Chinese, ii) Arabic and Vietnamese, and iii) Chinese and Vietnamese. Consistently, the SMI functions outperform other methods in terms of Avg. WER and TF score. Evidently, FLMI performs the best across all accent pairs. In \figref{fig:l2-multi-plot}, for the CHN-VTN accent pair, we demonstrate the Avg. WER improvement and duration of utterances selected from both the accents across a wide range of budgets. We observe that FLMI achieves the highest Avg. WER and selects the most number of utterances from both target accents, proving it can achieve robust targeting and fairness.




\section{Conclusion}
\label{sec:conclusion}
In this work, we propose \model, a data efficient and fair targeted subset selection method for ASR accent adaptation. \model\ utilizes submodular mutual information (SMI) functions to find representative speech utterances that belong to the target accent within a limit budget. We show that SMI functions consistently outperform other methods for targeting. We also demonstrate that \model\ is capable of targeting multiple accents fairly, which can be beneficial for deploying ASR models in regions with populations that speak in more than one accent.

\section{Limitations}
\label{sec:limit}
Similar to the limitations of existing selection methods, our method needs a reasonable feature embedding for accent representation in order to effectively target accents. MFCC features are not the best choice to represent accent information. Some accents may be more difficult to represent than others. This also lowers fairness scores for such accents. For instance, in one of our experiments where Manipuri accent was paired with Rajasthani or Assamese accents, we observe that acquiring a fair subset using any selection strategy is challenging (see \tabref{tab:IndicTTS-Combined-Appendix}). Although, FLMI was able to achieve a higher TF score than others, it was relatively lower than other accent pairs (see \tabref{tab:IndicTTS-2} and \tabref{tab:l2-2}). This is due to the fact that the pairwise similarity scores of utterances within the Manipuri accent are lower than other accents. The lower pairwise similarity scores lead to lower marginal gains during greedy maximization and are a consequence of poor feature representations due to insufficient information being encoded about the Manipuri accent. On another note, a risk associated with the targeting ability of \model\ is that it could be misused to create models that are unfair to certain populations. For future work, evaluating the performance of \model\ on larger datasets and other diverse settings (\eg\ out-of-distribution accents) will be interesting.



\section{Acknowledgments and Disclosure of Funding}
This work is supported by an Amazon Research Awarded (AWA) awarded to Preethi Jyothi, Ganesh Ramakrishnan and Rishabh Iyer, and by the National Science Foundation under Grant No. IIS-2106937 awarded to Rishabh Iyer. Any opinions, findings, and conclusions or recommendations expressed in this material are those of the authors and do not necessarily reflect the views of Amazon or the National Science Foundation.

\bibliography{acl2023}
\bibliographystyle{acl_natbib}








\end{document}

%% file: tables/indic/indic-1.tex
\setlength\tabcolsep{2pt}

\begin{table*}[ht]
\centering
\scalebox{0.95}{
\begin{tabular}{|c|c|c|c|c|c|c|c|c|}
\hline
\textbf{Eval} & \textbf{Pre} & \textbf{Random} & \textbf{Entropy} & \textbf{PhoneDiv} & \textbf{FL} & \textbf{LogDet} & \textbf{GCMI}   & \textbf{FLMI}  \\\hline
ASM ($\mathbf{T\%}$) & 27.1 & 21.5 {\small (16.7\%)} & 26.4 {\small (2.2\%)}  & 22.7 {\small (9.3\%)}  & 21.7 {\small (16.1\%)} & 22.2 {\small (13.6\%)} & \textbf{18.1} {\small (100\%)} & 18.7 {\small (100\%)}         \\
GUJ ($\mathbf{T\%}$) & 13.7 & 11.0 {\small (15.6\%)} & 11.2 {\small (28.8\%)} & 11.1 {\small (1.5\%)}  & 10.9 {\small (22.6\%)} & 10.7 {\small (24.6\%)} & 9.7 {\small (100\%)}  & \textbf{9.4} {\small (100\%)} \\
HIN ($\mathbf{T\%}$) & 11.1 & 9.7 {\small (9.8\%)}   & 9.7 {\small (11.0\%)}  & 10.4 {\small (1.7\%)}  & 9.5 {\small (10.1\%)}  & 9.7 {\small (12.1\%)}  & \textbf{8.5} {\small (100\%)}  & 9.2 {\small (100\%)}          \\
KAN ($\mathbf{T\%}$) & 18.7 & 15.3 {\small (9.8\%)}  & 15.4 {\small (14.4\%)} & 17.8 {\small (2.3\%)}  & 15.7 {\small (7.1\%)}  & 16.0 {\small (2.5\%)}  & \textbf{12.8} {\small (100\%)} & 13.1 {\small (100\%)}         \\
MAL ($\mathbf{T\%}$) & 19.5 & 16.8 {\small (12.2\%)} & 16.9 {\small (8.7\%)}  & 18.8 {\small (1.7\%)}  & 18.6 {\small (8.6\%)}  & 18.3 {\small (4.1\%)}  & 13.9 {\small (100\%)} & \textbf{13.6} {\small (98.6\%)} \\
MAN ($\mathbf{T\%}$) & 53.1 & 44.8 {\small (13.3\%)} & 48.4 {\small (5.3\%)}  & 42.5 {\small (79.9\%)} & 44.5 {\small (10.8\%)} & 43.3 {\small (20.6\%)} & \textbf{39.8} {\small (100\%)} & 39.9 {\small (100\%)}         \\
RAJ ($\mathbf{T\%}$) & 21.9 & 16.9 {\small (7.5\%)}  & 16.3 {\small (11.5\%)} & 18.2 {\small (1.9\%)}  & 16.9 {\small (9.4\%)}  & 16.4 {\small (9.4\%)}  & \textbf{14.3} {\small (100\%)} & 14.4 {\small (100\%)}         \\
TAM ($\mathbf{T\%}$) & 12.5 & 11.9 {\small (15.1\%)} & 11.7 {\small (18.0\%)} & 12.2 {\small (1.6\%)}  & 11.9 {\small (15.3\%)} & 11.9 {\small (13.0\%)} & \textbf{11.1} {\small (100\%)} & 11.5 {\small (100\%)}         \\\hline
\end{tabular}}
\caption{Single-target accent adaptation in IndicTTS on a 1230s budget.}
\label{tab:indic-1}
\end{table*}

%% file: tables/l2/l2-1.tex
\setlength\tabcolsep{2pt}

\begin{table*}[!ht]
\centering
\scalebox{0.95}{
\begin{tabular}{|c|c|c|c|c|c|c|c|c|}
\hline
\textbf{Eval} & \textbf{Pre} & \textbf{Random} & \textbf{Entropy} & \textbf{PhoneDiv} & \textbf{FL} & \textbf{LogDet} & \textbf{GCMI}   & \textbf{FLMI}  \\\hline
ARB ($\mathbf{T\%}$) & 24.3 & 23.1 {\small (14.7\%)} & 23.7 {\small (16.6\%)} & 23.3 {\small (15.2\%)} & 23.4 {\small (15.1\%)} & 23.4 {\small (8.1\%)}  & 22.8 {\small (58.6\%)}          & \textbf{20.8} {\small (98.7\%)} \\
CHN ($\mathbf{T\%}$) & 30.7 & 28.3 {\small (19.4\%)} & 28.5 {\small (18.1\%)} & 28.3 {\small (20.4\%)} & 28.2 {\small (16.8\%)} & 28.3 {\small (14.9\%)} & 25.8 {\small (70.4\%)}          & \textbf{25.1} {\small (99.6\%)} \\
HIN ($\mathbf{T\%}$) & 18.1 & 17.1 {\small (15.2\%)} & 18.6 {\small (7.7\%)}  & 16.4 {\small (14.1\%)} & 17.3 {\small (14.2\%)} & 16.5 {\small (28.7\%)} & 15.5 {\small (48.0\%)}          & \textbf{15.4} {\small (92.9\%)} \\
KOR ($\mathbf{T\%}$) & 19.1 & 17.7 {\small (17.5\%)} & 18.4 {\small (15.7\%)} & 18.2 {\small (16.0\%)} & 18.2 {\small (17.3\%)} & 17.6 {\small (23.5\%)} & 17.0 {\small (84.3\%)}          & \textbf{16.5} {\small (98.9\%)} \\
ESP ($\mathbf{T\%}$) & 23.4 & 22.2 {\small (16.6\%)} & 22.5 {\small (23.5\%)} & 22.4 {\small (17.2\%)} & 23.3 {\small (17.9\%)} & 22.4 {\small (7.0\%)}  & 21.3 {\small (89.2\%)}          & \textbf{20.8} {\small (99.6\%)} \\
VTN ($\mathbf{T\%}$) & 37   & 33.7 {\small (16.6\%)} & 35.5 {\small (18.4\%)} & 34.5 {\small (17.1\%)} & 34.2 {\small (18.6\%)} & 34.9 {\small (17.7\%)} & \textbf{31.2} {\small (97.1\%)} & 31.6 {\small (100\%)}        \\\hline
\end{tabular}}
\caption{Single-target accent adaptation in L2-Arctic on a 1230s budget.}
\label{tab:l2-1}
\end{table*}

%% file: tables/indic/Combined.tex
\setlength\tabcolsep{2pt}

\begin{table*}[!ht]
\centering
\scalebox{0.9}{
\begin{tabular}{|c|c|c|c|c|c|c|c|c|}
\hline
\textbf{Eval} & \textbf{Pre} & \textbf{Random} & \textbf{Entropy} & \textbf{PhoneDiv} & \textbf{FL} & \textbf{LogDet} & \textbf{GCMI}   & \textbf{FLMI}  \\\hline
{\small ASM-WER {\small ($\mathbf{T\%}$)}}          & 27.1 & 20.4 {\small (17.8\%)} & 24.8 {\small (1.9\%)} & 20.7 {\small (17.7\%)} & 20.1 {\small (16.2\%)} & 21.0 {\small (15.1\%)} & 21.1 {\small (24.7\%)}          & \textbf{19.7} {\small (50.6\%)} \\
{\small MAL-WER {\small ($\mathbf{T\%}$)}}           & 19.5 & 16.2 {\small (10.9\%)} & 16.8 {\small (8.6\%)} & 18.4 {\small (3.8\%)}  & 18.7 {\small (8.7\%)}  & 17.5 {\small (4.4\%)}  & \textbf{14.0} {\small (75.3\%)} & 14.7 {\small (49.4\%)}          \\
{\small Targeted Fairness (TF)} &      -& 0.08          & 0.01         & 0.03          & 0.06          & 0.03          & 0.74                   & \textbf{1}             \\
{\small Avg. WER}          &   23.3   & 18.3          & 20.8         & 19.5          & 19.4          & 19.2          & 17.6                   & \textbf{17.2}    \\
\hline
{\small MAL-WER ($\mathbf{T\%}$)}   & 19.5 & 16.2 {\small (10.9\%)} & 16.8 {\small (8.6\%)}  & 18.4 {\small (3.8\%)} & 18.7 {\small (8.7\%)} & 17.5 {\small (4.4\%)} & 15.7 {\small (7.6\%)}           & \textbf{15.0} {\small (21.9\%)} \\
{\small RAJ-WER ($\mathbf{T\%}$)}   & 21.9 & 16.2 {\small (8.5\%)}  & 15.8 {\small (12.4\%)} & 17.2 {\small (4.0\%)} & 15.9 {\small (8.9\%)} & 16.1 {\small (7.5\%)} & \textbf{13.3} {\small (92.4\%)} & 14.3 {\small (78.1\%)}          \\
{\small Targeted Fairness (TF)} &      -& 0.04          & 0.04          & 0.01         & 0.03         & 0.01         & 0.28                   & \textbf{0.68}          \\
{\small Avg. WER}  &   20.7   & 16.2          & 16.3          & 17.8         & 17.3         & 16.8         & \textbf{14.5}          & 14.6                   \\
\hline
{\small ASM-WER ($\mathbf{T\%}$)}   & 27.1 & 20.4 {\small (17.8\%)} & 24.8 {\small (1.9\%)}  & 20.7 {\small (17.7\%)} & 20.1 {\small (16.2\%)} & 21.0 {\small (15.1\%)} & 25.6 {\small (0.0\%)}            & \textbf{21.2} {\small (25.5\%)} \\
{\small RAJ-WER ($\mathbf{T\%}$)}   & 21.9 & 16.2 {\small (8.5\%)}  & 15.8 {\small (12.4\%)} & 17.2 {\small (4.0\%)}  & 15.9 {\small (8.9\%)}  & 16.1 {\small (7.5\%)}  & \textbf{13.3} {\small (100\%)} & 14.3 {\small (74.5\%)}          \\
{\small Targeted Fairness (TF)} &      -& 0.06          & 0.01          & 0.03          & 0.06          & 0.05          & 0                       & \textbf{0.76}          \\
{\small Avg. WER}          &   24.5   & 18.3          & 20.3          & 19            & 18            & 18.6          & 19.5           & \textbf{17.8}          \\
\hline
\end{tabular}
}
\caption{Targeting Assamese, Rajasthani and Malayalam accents in pairs on a 2460s budget (IndicTTS). For definitions of $\mathbf{T\%}$ and Targeted Fairness (TF), please refer to \secref{sec:single-acc} and \secref{sec:two-acc}.}
\label{tab:IndicTTS-2}
\end{table*}


%% file: tables/l2/Combined.tex
\setlength\tabcolsep{2pt}

\begin{table*}[!ht]
\centering
\scalebox{0.9}{
\begin{tabular}{|c|c|c|c|c|c|c|c|c|}
\hline
\textbf{Eval} & \textbf{Pre} & \textbf{Random} & \textbf{Entropy} & \textbf{PhoneDiv} & \textbf{FL} & \textbf{LogDet} & \textbf{GCMI}   & \textbf{FLMI}  \\\hline
{\small ARB-WER ($\mathbf{T\%}$)}   & 24.3 & 22.9 {\small (15.4\%)} & 23.4 {\small (17.2\%)} & 22.5 {\small (14.1\%)} & 23.4 {\small (15.2\%)} & 22.9 {\small (9.9\%)}  & 21.0 {\small (44.3\%)} & \textbf{20.3} {\small (63.5\%)} \\
{\small CHN-WER ($\mathbf{T\%}$)}   & 30.7 & 27.1 {\small (17.5\%)} & 28.3 {\small (19.0\%)} & 27.7 {\small (19.6\%)} & 27.5 {\small (17.2\%)} & 27.8 {\small (13.9\%)} & 26.3 {\small (31.5\%)} & \textbf{26.1} {\small (32.7\%)} \\
{\small Targeted Fairness (TF)} &      -& 0.11          & 0.13          & 0.11          & 0.1           & 0.06          & 0.56          & \textbf{0.83}          \\
{\small Avg. WER}          &   27.5   & 25            & 25.8          & 25.1          & 25.4          & 25.4          & 23.6          & \textbf{23.2}         \\
\hline
{\small CHN-WER ($\mathbf{T\%}$)}   & 30.7 & 27.1 {\small (17.5\%)} & 28.3 {\small (19.0\%)} & 27.7 {\small (19.6\%)} & 27.5 {\small (17.2\%)} & 27.8 {\small (13.9\%)} & 25.9 {\small (30.3\%)} & \textbf{25.3} {\small (67.3\%)} \\
{\small VTN-WER ($\mathbf{T\%}$)}   & 37   & 32.8 {\small (16.8\%)} & 34.4 {\small (16.2\%)} & 33.6 {\small (17.3\%)} & 33.2 {\small (17.5\%)} & 33.7 {\small (15.5\%)} & 33.9 {\small (15.5\%)} & \textbf{32.2} {\small (31.2\%)} \\
{\small Targeted Fairness (TF)} &      -& 0.12          & 0.12          & 0.14          & 0.12          & 0.09          & 0.19          & \textbf{0.84}          \\
{\small Avg. WER}          &   33.8   & 30            & 31.4          & 30.6          & 30.4          & 30.8          & 29.9          & \textbf{28.8}         \\
\hline
{\small ARB-WER ($\mathbf{T\%}$)}   & 24.3 & 22.9 {\small (15.4\%)} & 23.4 {\small (17.2\%)} & 22.5 {\small (14.1\%)} & 23.4 {\small (15.2\%)} & 22.9 {\small (9.9\%)}  & 21.3 {\small (40.8\%)}          & \textbf{20.3} {\small (74.2\%)} \\
{\small VTN-WER ($\mathbf{T\%}$)}   & 37   & 32.8 {\small (16.8\%)} & 34.4 {\small (16.2\%)} & 33.6 {\small (17.3\%)} & 33.2 {\small (17.5\%)} & 33.7 {\small (15.5\%)} & \textbf{33.8} {\small (20.8\%)} & 34.1 {\small (19.9\%)}          \\
{\small Targeted Fairness (TF)} &      -& 0.1           & 0.11          & 0.1           & 0.11          & 0.06          & 0.34                   & \textbf{0.59}          \\
Avg. WER          &   30.6   & 27.8          & 28.9          & 28            & 28.3          & 28.3          & 27.6                   & \textbf{27.2}         \\
\hline
\end{tabular}
}
\caption{Targeting Vietnamese, Arabic and Chinese accents in pairs on a 2460s budget (L2-Arctic).}

\label{tab:l2-2}
\end{table*}

%% file: tables/indic/Combined-Appendix.tex
\setlength\tabcolsep{2pt}

\begin{table*}[ht]
\centering
\scalebox{0.9}{
\begin{tabular}{|c|c|c|c|c|c|c|c|c|}
\hline
\textbf{Eval} & \textbf{Pre} & \textbf{Random} & \textbf{Entropy} & \textbf{PhoneDiv} & \textbf{FL} & \textbf{LogDet} & \textbf{GCMI}   & \textbf{FLMI}  \\\hline
{\small ASM-WER {\small ($\mathbf{T\%}$)}}   & 27.1 & 20.4 {\small (17.8\%)} & 24.8 {\small (1.9\%)} & 20.7 {\small (17.7\%)} & 20.1 {\small (16.2\%)} & 21.0 {\small (15.1\%)} & 17.1 {\small (97.8\%)} & \textbf{17.0} {\small (92.6\%)} \\
{\small MAN-WER {\small ($\mathbf{T\%}$)}}   & 53.1 & 42.5 {\small (13.6\%)} & 47.2 {\small (3.8\%)} & 40.7 {\small (59.0\%)} & 42.4 {\small (12.9\%)} & 40.7 {\small (23.9\%)} & 44.6 {\small (2.2\%)}  & \textbf{42.9} {\small (7.4\%)}  \\
{\small Targeted Fairness (TF)} &      -& 0.1           & 0            & 0.42          & 0.08          & 0.14          & 0.09          & \textbf{0.27}          \\
{\small Avg. WER}          &   40.1   & 31.4          & 36           & 30.7          & 31.2          & 30.8          & 30.8          & \textbf{30}            \\
\hline
{\small MAN-WER ($\mathbf{T\%}$)}   & 53.1 & 42.5 {\small (13.6\%)} & 47.2 {\small (3.8\%)}  & 40.7 {\small (59.0\%)} & 42.4 {\small (12.9\%)} & 40.7 {\small (23.9\%)} & 48.9 {\small (0.0\%)}            & \textbf{47.2 {\small (2.7\%)}} \\
{\small RAJ-WER ($\mathbf{T\%}$)}   & 21.9 & 16.2 {\small (8.5\%)}  & 15.8 {\small (12.4\%)} & 17.2 {\small (4.0\%)}  & 15.9 {\small (8.9\%)}  & 16.1 {\small (7.5\%)}  & \textbf{13.2 {\small (100\%)}} & 13.7 {\small (97.3\%)}         \\
{\small Targeted Fairness (TF)} &      -& 0.05          & 0.02          & 0.09          & 0.05          & 0.07          & 0                       & \textbf{0.11}         \\
{\small Avg. WER}          &  37.5    & 29.4          & 31.5          & 29            & 29.2          & 28.4          & 31                      & \textbf{30.4}         \\
\hline
\end{tabular}
}
\caption{Targeting Assamese, Rajasthani and Manipuri accents in pairs on a 2460s budget (IndicTTS)}
\label{tab:IndicTTS-Combined-Appendix}
\end{table*}